# Optical Design of the SuMIRe PFS Spectrograph

Sandrine Pascal*[a], Sébastien Vives [a], Robert H. Barkhouser [b], James E. Gunn [c]

[a] Aix Marseille Université - CNRS, LAM (Laboratoire d'Astrophysique de Marseille) , UMR 7326, 13388, Marseille, France;

[b] Johns Hopkins University, Department of Physics and Astronomy, 3701 San Martin Drive, Baltimore, MD 21218, USA;

[c] Princeton University, Department of Astrophysical Sciences, Princeton, NJ 08544, USA;

## ABSTRACT

The SuMIRe Prime Focus Spectrograph (PFS), developed for the 8-m class SUBARU telescope, will consist of four identical spectrographs, each receiving 600 fibers from a 2394 fiber robotic positioner at the telescope prime focus. Each spectrograph includes three spectral channels to cover the wavelength range [0.38-1.26] um with a resolving power ranging between 2000 and 4000. A medium resolution mode is also implemented to reach a resolving power of 5000 at 0.8 um.

Each spectrograph is made of 4 optical units: the entrance unit which produces three corrected collimated beams and three camera units (one per spectral channel: "blue, "red", and "NIR"). The beam is split by using two large dichroics; and in each arm, the light is dispersed by large VPH gratings (about 280x280mm).

The proposed optical design was optimized to achieve the requested image quality while simplifying the manufacturing of the whole optical system. The camera design consists in an innovative Schmidt camera observing a large field-of-view (10 degrees) with a very fast beam (F/1.09). To achieve such a performance, the classical spherical mirror is replaced by a catadioptric mirror (i.e meniscus lens with a reflective surface on the rear side of the glass, like a Mangin mirror).

This article focuses on the optical architecture of the PFS spectrograph and the performance achieved. We will first described the global optical design of the spectrograph. Then, we will focus on the Mangin-Schmidt camera design. The analysis of the optical performance and the results obtained are presented in the last section.

## 1. INTRODUCTION

The Prime Focus Spectrograph (PFS) of the Subaru Measurement of Images and Redshifts (SuMIRe) project is developed for the SUBARU telescope at Mauna Kea, Hawaii. This optical and near-infrared multi-fiber spectrograph targets cosmology with galaxy surveys, galactic archaeology, and studies of galaxy/AGN evolution [4].

The PFS spectrograph is composed of four identical modules (spectrograph), each one receiving 600 fibers from a 2394 fiber robotic positioner at the telescope prime focus. To meet the scientific requirements, the spectrographs have to deal with a large field of view (10 degrees) and a very fast f-ratio (F/1.09). Furthermore, the fabrication of four identical modules enhance the need to facilitate the manufacturability. The optical architecture of each spectrograph is based on a Schmidt collimator facing a Schmidt camera. The camera concept was slightly modified, replacing the classical spherical mirror by a Mangin-like mirror (i.e meniscus lens with a reflective surface on the rear side of the glass).

*contact: sandrine.pascal@lam.fr; phone 33 495044147; www.lam.fr

## 2. SPECTROGRAPH MODULE OPTICAL DESIGN

The optical concept of one spectrograph module is presented in Fig 1. The main optical sub-systems of a spectrograph module are:

- the entrance slit which consists of about 600 fibers arranged on a curved slit;

- the entrance unit which produces three corrected collimated beams and disperses light;

- the three camera units (one per spectral channel: "blue, "red", and "NIR");

The wavebands and resolving powers corresponding to each spectral channel are presented in Fig. 2. The red channel has two dispersive elements mounted on an exchange mechanism to provide a medium resolution observing mode in addition of the low resolution mode.

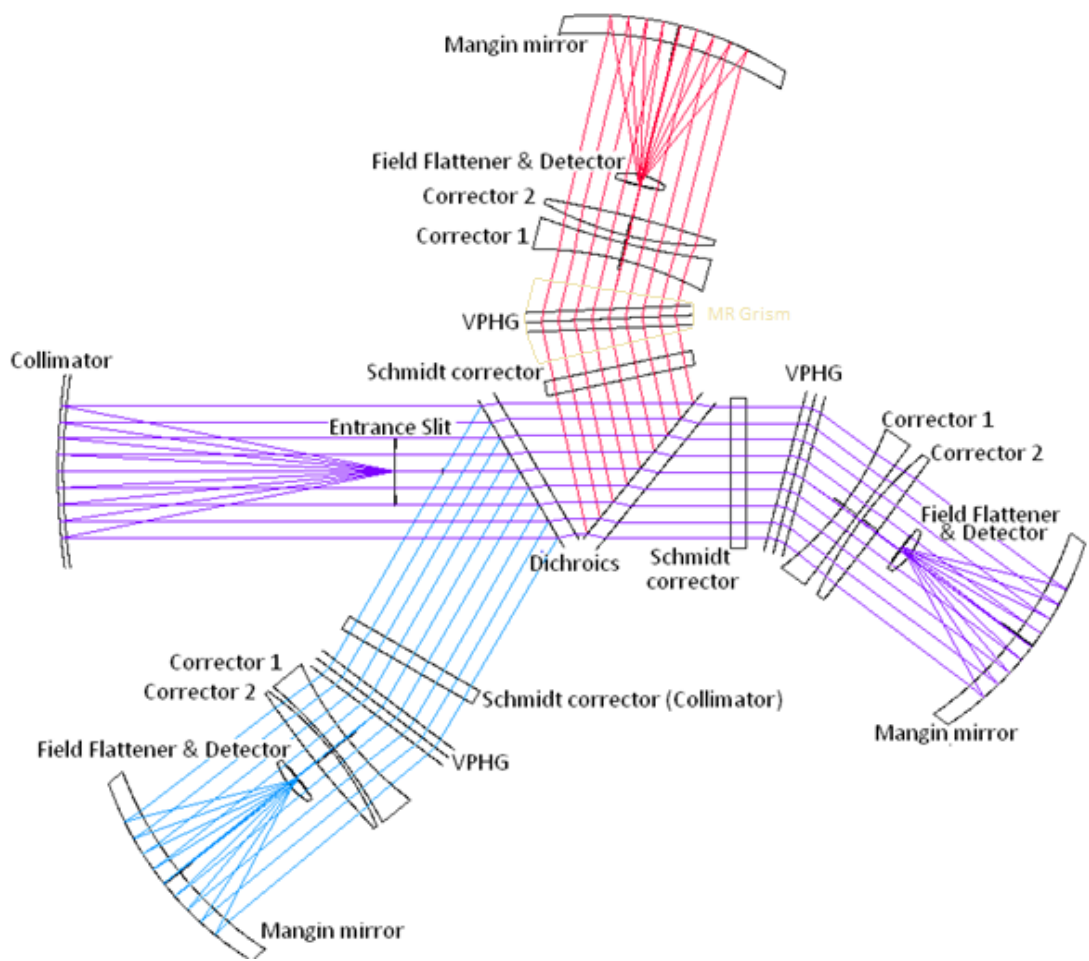

Figure 1.Spectrograph Module Layout

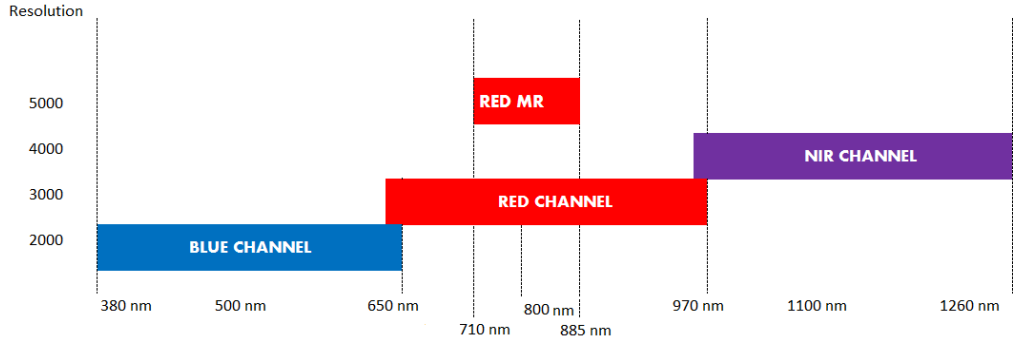

Figure 2. Bandwidths and resolution of PFS Spectrograph Channels.

### 2.1 Pseudo-Slit

The spectrograph object consists of about 600 individual optical fibers coming from the primary focus. The fibers (128 um core diameter) are arranged on a curved slit with a slit height of 69.4 mm. This results in a typical pitch of 255 um between fibers.

The nominal output f-ratio of the fibers is 2.8 but the focal ratio degradation (FRD) can decrease this f-ratio. The FRD expected in the fibers is not strongly well known at this time but we assumed to have some light between 2.8 and 2.5. In order to avoid light losses due to vignetting of the beam beyond 2.8, we considered a f-ratio of 2.55 for the optical design.

A cover plate of about 100 um thick will be placed in front of the slit to limit the amount of unwanted reflections on the slit.

### 2.2 Entrance Unit

The Entrance Unit delivers three collimated beams using a collimator mirror and three identical Schmidt correctors (one by spectral channel). The wavebands are separated by two plates with dichroics coatings manufactured by ASAHI (Japan).

In the spectrograph, the light is dispersed by three VPH gratings manufactured by Kaiser Optical Systems Inc. (USA) and one VPH grism for the medium resolution mode of the red channel (see section 2). A detailed description of the VPH gratings designs and performance tests is given in [1].

### 2.3 Camera Unit

In each of the three channels, the camera images the spectra on a square 4K x 4K detector with a pixel size of 15 um. Each camera is based on a Mangin-Schmidt concept adapted to the fast focal ratio (F/1.09) and to the large field of view.

In order to simplify the manufacturing process of the four spectrograph, we used identical Mangin mirrors for the blue and the red channel. The optical design of the camera is studied in the section 3.

## 3. MEDIUM RESOLUTION MODE

A medium-resolution mode, with resolving power of R=5000, was implemented on the red channel in addition of the low-resolution mode (see Fig. 3). This MR mode will enable to investigate further important aspects in the galactic archeology science case.

The covered wavelength range is 1750 A wide and reach from 7100 A to 8850 A. The red VPH grating and the MR-mode can be exchanged without having to change the position of the camera. Indeed, two prisms are attached to the VPH surfaces to keep the camera location fixed between the two resolution mode. The two prisms have an angle of 13°, they are made in OHARA-LAH53 glass in order to meet the resolving power specification. The central VPH grating have a groove spacing of 1007 lines/um in order to center the wavelength of 7992 A on the detector.

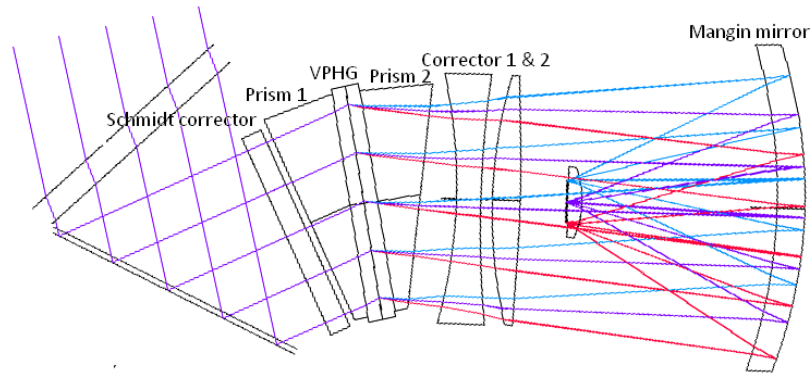

Figure 3. Medium Resolution Mode Layout for Red channel.

## 4. MANGIN-SCHMIDT CAMERA

The Mangin-Schmidt camera design was selected to achieve the fast focal f-ratio f/1.09 with the required image quality over the FoV. Each camera is composed of the following elements (see Fig 4):

- A double Schmidt corrector which consists in two silica lenses with one aspherical surface per corrector (correctors are different for each channel);

- A spherical Mangin-like mirror. It consists in a meniscus lens of 40 mm thick made in silica glass with a reflective surface on the rear face. The front and rear faces are spherical with relatively similar curvatures. (Red and Blue Mangin mirrors are identical, NIR is different)

- A field flattener lens made in Silica with the front and the rear surfaces being aspherical (field flattener are different for each channel).

A configuration with a double corrector was chosen to limit the deformation of a single Schmidt corrector. Indeed, with only one corrector, the maximal deformation of each face from the best sphere was ranging between 2.2 and 3.9 mm (require specific CGH developments for inspection) while the use of two correctors allows lower deformations. This solution was selected to reduce manufacturing difficulties and limit the risks.

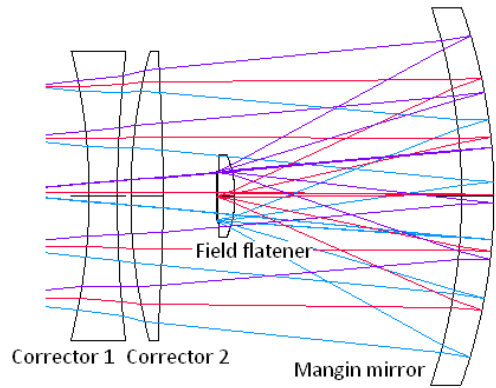

Figure 5: Mangin-Schmidt camera layout and optical components.

**4.1 Comparison between Mangin-Schmidt and Schmidt configuration**

A comparison with the standard Schmidt camera shows that the use of the Mangin-Schmidt leads to a design with simpler optical elements (Schmidt corrector especially) and better image quality.

The typical improvement of image quality considering the average RMS Spot radius (over fields and wavelengths) is about 8%. The results obtained show that the use of the glass on the mirror is a good solution to improve image quality while keeping the manufacturing easy.

However, one should mention that the use of a Mangin mirror induces inherent ghosts relative to unwanted reflections on the glass. Compared to the simple mirror, we have to consider the possible reflection on the front face before entering the glass and the multiple reflections in the glass (especially the double reflection). In the spectrograph camera, this results in two out-of-focus ghosts which cover roughly the entire detector area. The contribution of these two ghosts is just under 2 times the anti-reflective coating reflectivity of the Mangin glass (1% in our case) and represent an important part of the ghost level in the camera.

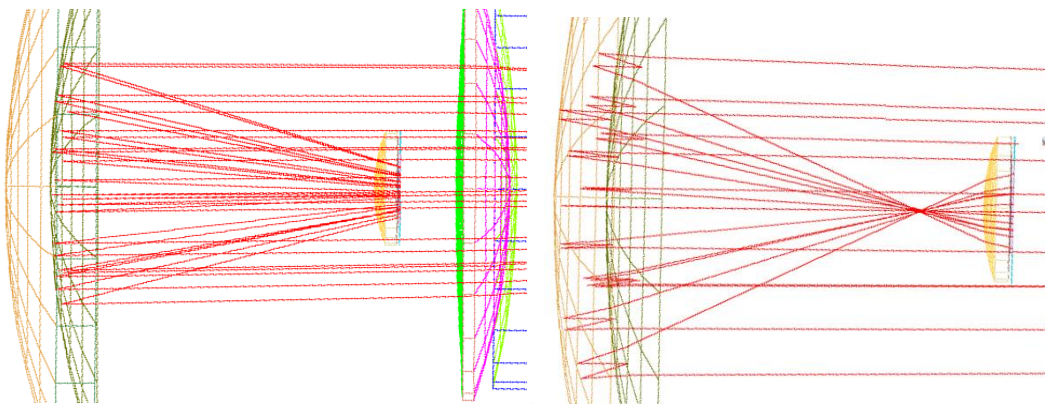

Figure 6: Layout of the Mangin ghosts: Reflection on the front face (left) and double reflection (right)

Another drawback concerns the tolerancing of the camera: the radii of curvature of the Mangin mirror for both faces are more sensitive than those of the usual Schmidt mirror. Consequently the tolerance given to Mangin radii were tighten by a factor two with respect to classical mirror in order to maintain image quality performance.

## 5. PERFORMANCE

### 5.1 Image Quality Specification

The requirement on image quality for the spectrograph is defined in terms of Ensquared Energy in a 3 by 3 pixels and Ensquared Energy in a 5 by 5 pixels as below:

The Ensquared Energy (EE) for a spatial element (fiber) shall be:
- ≥ 50% within a square of 3 pixels for each spectral band in more than 95% of the detector area;
- ≥ 90% within a square of 5 pixels for each spectral band in more than 95% of the detector area.

The estimation of this criterion takes into account many contributors, especially the convolution with the fiber core diameter. The best way to model the convolution with the fiber in the optical design is to use Extended Ensquared Energy but it is time consuming to compute the EEE during optimization and analysis of the optical design. To handle that, we tried to find a relationship/correlation between standard image quality criteria such as RMS spot radius (or geometric ensquared energy) and Extended Ensquared Energy but no strong correlation was found. Fig 7. presents the relationship between RMS spot radius and Extended Ensquared Energy for a fiber of 128 um core. We can see that for one value of EEE corresponds a large range of RMS spot radius.

For these reasons, we finally consider usual criteria but also Extended Ensquared Energy during the optimization and analysis of the optical design.

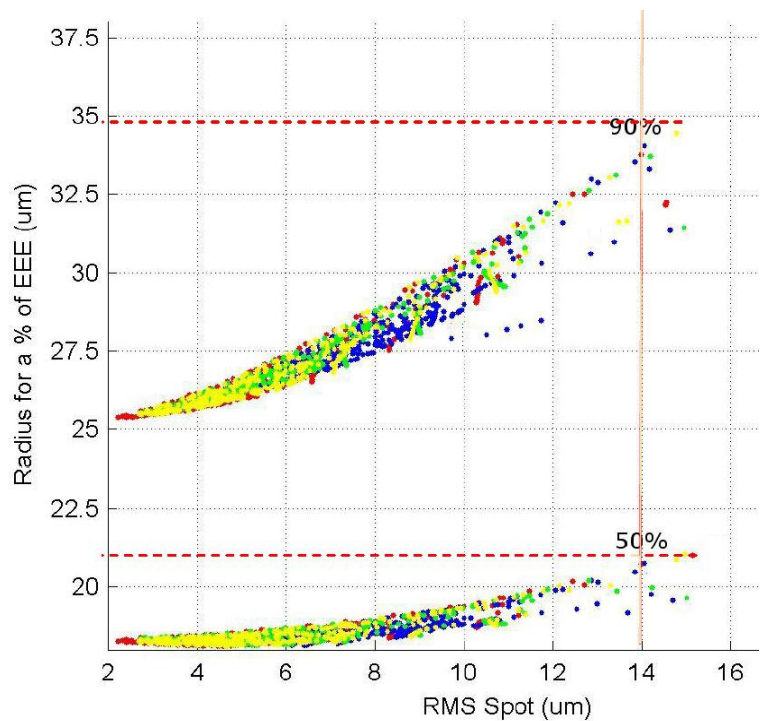

Figure 8: Relation between EEE and RMS spot radius on toleranced design. One point represents a couple FoV/wavelengths, colors are for slighlty different designs.

## 5.2 Matrix Spot Diagram

Fig 9, 10, and 11 shows the matrix spot diagram of the nominal design for each channel (blue, red, NIR):

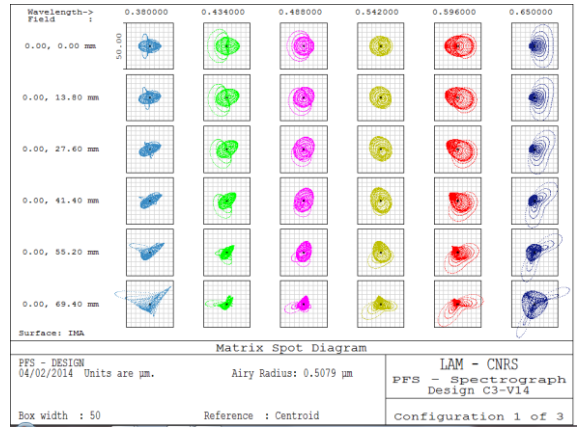

Figure 12: Matrix Spot Diagram Blue Channel (box size is 50 um)

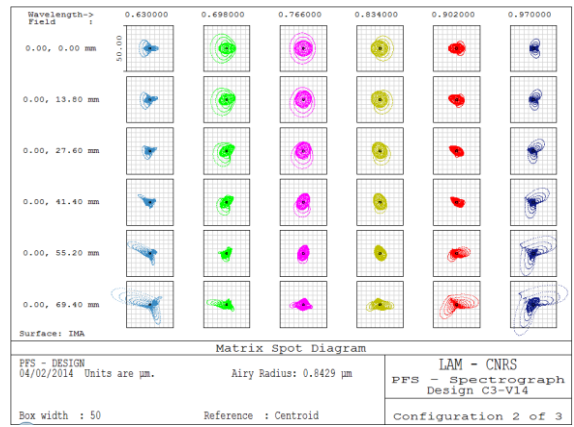

Figure 13: Matrix Spot Diagram Red Channel (box size is 50 um)

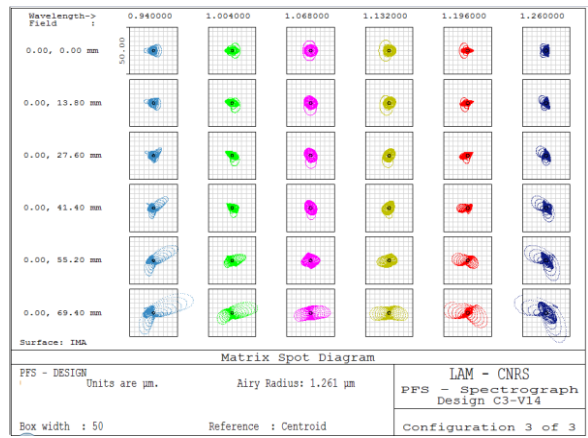

Figure 14: Matrix Spot Diagram NIR Channel (box size is 50 um)

## 5.3 Performance obtained on toleranced design

The tolerances on the main optical components were defined with a bottom-up approach: We first defined what was reasonably achievable, then we modify and refine these tolerances according to the performance needed and obtained. This results in the final tolerances after a few iterations. The tolerances adopted this way allow to reach the image quality required without major difficulties in the manufacturing process. Note that tolerances on VPH gratings are not taken into account here and discussed in [1].

The following figures show the distribution of the square size (in pixels) containing respectively 50% and 90% of EEE for the nominal design (yellow) and for a set of toleranced designs (blue), considering manufacturing and alignment tolerances of nearly all the optical components. This is done with a fiber of 60 um corresponding to the "engineering" fibers (also used for image quality verifications). Each couple FoV/wavelenght is represented (no mean over field or wavelength). The results show that the mean value of the distribution for the toleranced designs increase just a bit but some couples FoV/wavelengths can possibly be much degraded.

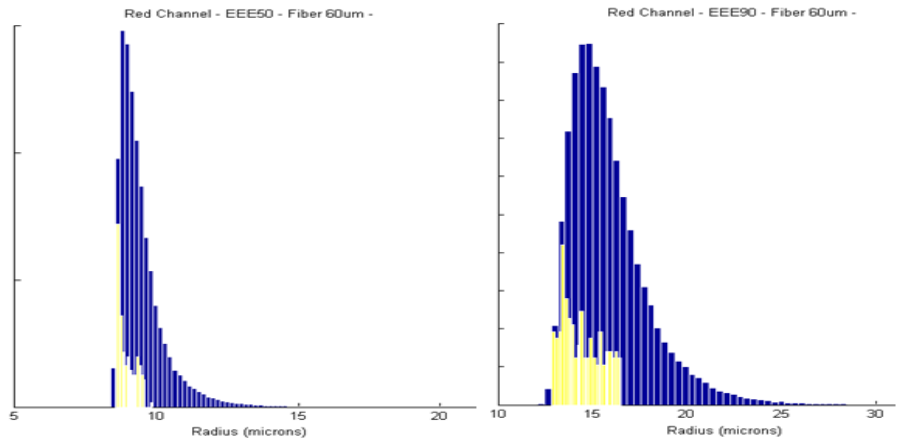

Figure 15: Distribution of the Extended Ensquared Energy for different detector positions (FoV / wavelengths) on the red channel. Fiber is 60 um core diameter. Yellow is for nominal design and blue distribution is for the toleranced designs.

## CONCLUSION

The concept based on a Schmidt collimator facing a Mangin-Schmidt camera presented in this paper allows us to reach the high image quality needed with few simple elements at the expense of the central obscuration, which leads to larger optics. The optical design has already passed the critical design review and optical components are now under manufacturing. The first spectrograph module will be integrated in 2015.

## ACKNOWLEDGMENTS

We acknowledge support from the Funding Program for World-Leading Innovative R&D in Science and Technology (FIRST), program: "Subaru Measurements of Images and Redshifts (SuMIRe)", CSTP, Japan.